\documentclass[reprint,amsmath,amssymb,aps,pra,longbibliography]{revtex4-2}
\usepackage{graphicx}
\usepackage{bm}
\usepackage{color, soul}
\usepackage{wasysym}
\usepackage{subfigure}
\usepackage{commath}
\usepackage{braket}
\usepackage{amsmath}
\usepackage{amsfonts}
\usepackage{amssymb}
\usepackage{amsthm}
\usepackage{siunitx}
\usepackage{multirow}
\usepackage{epstopdf}
\usepackage{array}
\usepackage{latexsym}
\usepackage{textcomp}
\usepackage{ulem}
\usepackage{braket}
\usepackage{siunitx}
\DeclareSIUnit\dBm{dBm}
\usepackage{booktabs}

\usepackage[title]{appendix}

\usepackage{float}
\usepackage{array}

\begin{document}

\title{Microcell CPT atomic clock using laser current-actuated power modulation\\ with 10$^{-12}$ range stability at 1 day}

\author{Carlos Manuel Rivera-Aguilar$^1$}
\thanks{Corresponding author: carlos.rivera@femto-st.fr}
\author{Andrei Mursa$^1$}
\author{Cl\'ement Carl\'e$^1$}
\author{Jean-Michel Friedt$^1$}
\author{Emmanuel Klinger$^1$}
\author{Moustafa Abdel Hafiz$^1$}
\author{Nicolas Passilly$^1$}
\author{Rodolphe Boudot$^1$}
\affiliation{$^1$ FEMTO-ST, CNRS, Universit\'e Marie et Louis Pasteur, SupMicrotech-ENSMM, Besan\c con, France.}

\date{\today}

\begin{abstract}
We present a coherent-population trapping (CPT) microcell atomic clock using symmetric auto-balanced Ramsey (SABR) spectroscopy. The pulsed SABR sequence is applied through direct current-based power modulation of the vertical-cavity surface-emitting laser, eliminating the need for an external optical shutter and enabling compatibility with fully-integrated clocks. The sequence is controlled by a single FPGA-based digital electronics board. A key aspect of proper clock operation was the implementation of a real-time tracking of the atomic signal detection window. The clock frequency dependence on laser power, microwave power, laser frequency, and timing of the detection window has been measured, obtaining sensitivity coefficients lower than those obtained with Ramsey-CPT spectroscopy. The Allan deviation of the SABR-CPT clock, based on a microfabricated cell with low-permeation glass windows, is 1.1~$\times$~10$^{-9}$ at 1~s and averages down to the low 10$^{-12}$ range at 1 day integration time. These results pave the way towards the development of Ramsey-CPT chip-scale atomic clocks with enhanced timekeeping performances.
\end{abstract}

\maketitle
\section{Introduction}
\label{sec:introduction}
Chip-scale atomic clocks (CSACs) based on coherent population trapping (CPT) have met a remarkable success in navigation, communication, defense and synchronization systems by offering a fractional frequency stability at 1 day in the 10$^{-11}$ range in a low size, weight and power (SWaP) budget \cite{Kitching:APR:2018}. These clocks rely on the interaction of a hot alkali vapor, confined in a microfabricated vapor cell in the presence of a pressure of buffer gas \cite{Dicke_1953}, with an optically-carried microwave signal obtained by direct current modulation of a vertical-cavity surface emitting laser (VCSEL). Under null Raman detuning condition, atoms are trapped in a quantum dark state \cite{Alzetta_1976} for which the atomic vapor transparency is increased, leading to the detection of an atomic resonance, used to stabilize the frequency of the local oscillator (LO) that drives the VCSEL.

The fractional frequency stability of CSACs is usually degraded for integration times higher than 100~s by light-shifts. Several approaches have been proposed to mitigate light-shifts in CPT-based CSACs. A widely used technique involves tuning the microwave power set point to cancel at the first order the clock frequency dependence to laser power variations  \cite{ZhuCutler:2000, Shah:APL:2006, Zhang:JOSAB:2016}. Nevertheless, this microwave power setpoint is specific to the physics-package, may not exist in cells filled with high buffer gas pressures \cite{Vaskovskaya:OE:2019}, and only protects the clock frequency from laser power variations. Other techniques, such as the implementation of advanced algorithms for compensation of the laser current-temperature couple \cite{Yanagimachi:APL:2020}, the tuning of the cell temperature \cite{Miletic:APB:2012}, or the use of interrogation sequences based on power-modulation \cite{Yudin:2020, MAH:PRap:2020, Yudin:2021}, have been proposed. 

Ramsey spectroscopy is efficient for light-shift reduction. Ramsey-CPT spectroscopy \cite{Thomas:PRL:1982, Zanon_PRL_2005_Ramsey}, in which atoms interact with a sequence of optical CPT pulses separated by a free-evolution dark time of length $T$, was performed first in microfabricated cells in Refs \cite{Boudot:JOSAB:2018, Carle:UFFC:2021}. The symmetric auto-balanced Ramsey (SABR) sequence \cite{MAH:APL:2018} was later demonstrated to reduce further the clock frequency dependence to laser power, laser frequency and microwave power \cite{MAH:2022}. This approach, combined with the use of a microcell made with low permeation glass substrates \cite{Carle:JAP:2023}, has enabled the demonstration of a microcell CPT clock with fractional frequency stability in the low 10$^{-12}$ range at 1 day \cite{Carle_OE_2023}.

Nevertheless, in the studies mentioned above, the pulsed optical sequence was produced using an external optical shutter, typically an acousto-optic modulator (AOM), which is incompatible with a fully-integrated CSAC. To address this challenge, operation of a Ramsey-CPT microcell atomic clock, with driving-current based power modulation of the VCSEL through a two-step pulse sequence \cite{Ide:IFCS:2015, Fukuoka:JJAP:2023, Fukuoka:JJAP:2024}, was recently reported \cite{Rivera:APL:2024}. This paper showed a tenfold reduction of the clock frequency sensitivity to laser power compared to the continuous-wave (CW) case. The clock stability, extracted from a 10-hour measurement, reached 1.3~$\times$~10$^{-10}$~$\tau^{-1/2}$, with limitations after 2000~s likely due to light-shifts and buffer gas permeation through the cell windows.
\begin{figure*}[t!]
    \centering
    \includegraphics[width=1.0\textwidth]{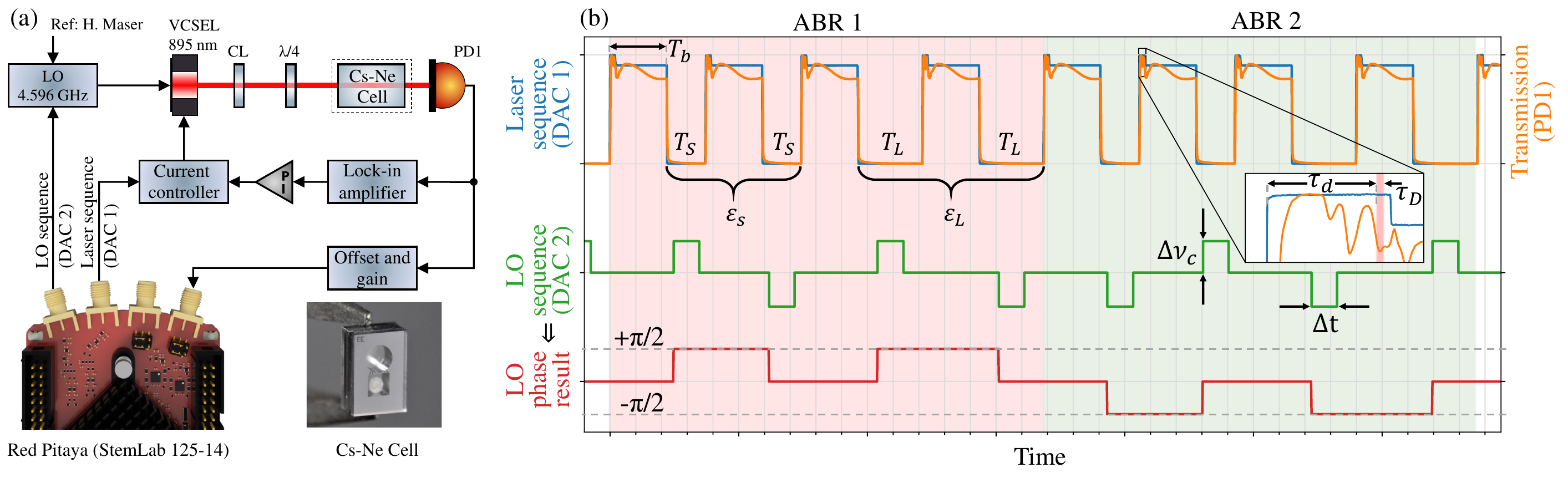}
    \caption{(a) Experimental setup. A hot Cs vapor, confined in a microfabricated vapor cell filled with a buffer gas pressure, interacts with a dual-frequency optical field obtained by direct modulation at 4.596 GHz of the VCSEL current. Atoms are trapped in the CPT dark state using both first-order optical sidebands.
    The signal at the cell output is used in three servo loops, for laser frequency and LO frequency, as well as light-shift compensation. CL: collimation lens LO: local oscillator. The inset shows a photograph of the Cs-Ne vapor microcell. The experiment is controlled by a single FPGA RedPitaya digital electronics board, shown in the lower-left corner. (b) SABR sequence used for interrogation of the atoms. Instead of interacting continuously with light as in traditional CPT clocks, atoms experience here a sequence of two-step pulses, of total length $T_b$, separated by free-evolution times ($T_S$ or $T_L$) in the dark, during which the light is actually turned off by tuning the laser current below the laser threshold current. The pulsed light sequence is then obtained by current-actuated power modulation of the laser.}
    \label{fig:1}
\end{figure*}

In this work, we demonstrate a Cs microcell CPT clock based on the SABR interrogation, which offers advantages over the Ramsey-CPT one. The SABR sequence, generated using an FPGA-based electronics board, is produced by directly modulating the VCSEL dc current, eliminating the need for an AOM and then being compliant with a fully-integrated CSAC. In this work, atoms in the microcell experience 170-$\mu$s-long optical CPT pulses, separated by a short dark time $T_s$ of 150~$\mu$s or a long dark time $T_L$ of 250~$\mu$s. The atomic signal is typically measured 18 $\mu$s after the beginning of each pulse in a 1~$\mu$s-long averaging window. An important step for proper operation of the clock was the implementation of a detection window tracking system, aimed to compensate for the timing jitter of the detection window caused by current step-induced thermal transients of the laser frequency along the sequence. The clock frequency sensitivity coefficients to laser power, microwave power, and laser frequency are all reduced in comparison with those obtained with Ramsey-CPT spectroscopy. The frequency stability of the microcell SABR-CPT clock, extracted from a 5-day measurement, is 1.1~$\times$~10$^{-9}$ at 1~s and reaches the level of 2.4~$\times$~10$^{-12}$ at 2~$\times$~10$^5$~s.
\begin{figure*}[t!]
    \centering
    \includegraphics[width=1.0\textwidth]{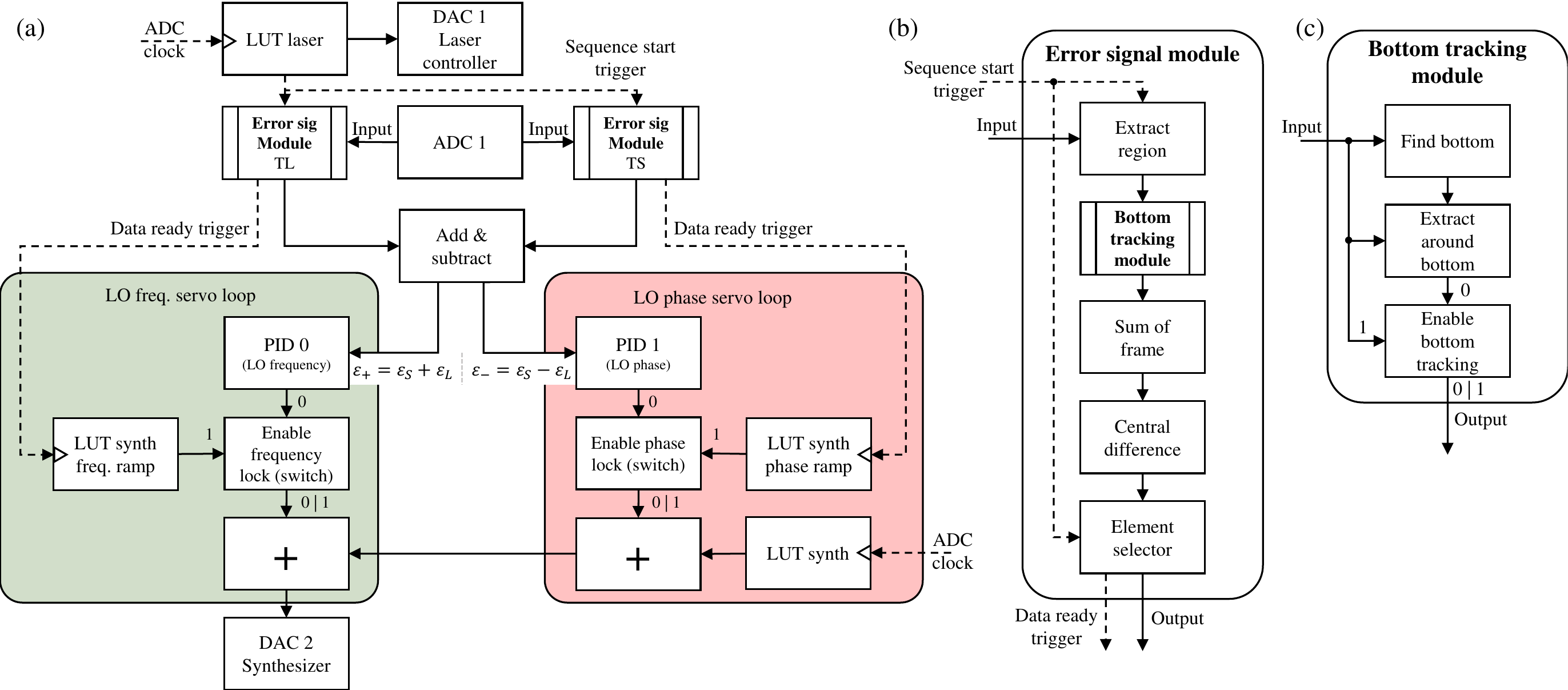}
    \caption{FPGA implementation of the SABR-CPT clock operation.}
    \label{fig:2}
\end{figure*}

\section{Experimental setup}
\label{sec:setup}
The experimental setup, shown in Fig. \ref{fig:1}(a), is comparable to the one described in Ref. \cite{Rivera:APL:2024}. The laser source is a VCSEL tuned on the Cs D$_1$ line at 895~nm \cite{Kroemer:AO:2016}. The laser is driven at 4.596~GHz by a commercial microwave synthesizer. The latter delivers a microwave power $P_{\mu W}$ of $-$5.5 dBm (282 $\mu$W) and is referenced to a hydrogen maser for frequency stability and frequency shift measurements. A lens is placed at the output of the laser to collimate the beam. A quarter-wave plate is placed for circularly polarizing the laser beam. The laser light, with a power $P_l$ of 188 $\mu$W, is sent into a physics package at the center of which is inserted a microfabricated Cs vapor cell filled with about 73~Torr (at 0$^{\circ}$C) of Ne buffer gas. The cell is stabilized at 82$^{\circ}$C. The atom-light interaction takes places in a 2-mm diameter and 1.5-mm long cavity etched in silicon, and filled with alkali vapor through post-sealing laser activation of a pill dispenser \cite{Hasegawa:SA:2011, Vicarini:SA:2018}. The cell is built with 500-$\mu$m-thick alumino-silicate glass (ASG) windows for reducing buffer gas permeation \cite{Carle:JAP:2023, Carle_OE_2023}.  A static magnetic field of 250~mG is applied to isolate the clock transition. A single-layer mu-metal magnetic shield protects the atoms from magnetic perturbations. At the output of the cell, the transmitted light is detected by a photodiode that delivers the atomic signal. Three servo loops are then implemented. The first one aims to stabilize the frequency of the laser onto the bottom of the absorption profile of interest. The laser frequency is connected to the $F\rq$~=~4 excited state. For this purpose, the laser current is modulated at a frequency of 13.3~kHz while the absorption signal at the photodiode output is demodulated with an external lock-in amplifier for generation of a zero-crossing error signal. When processed with a PI controller, this signal is exploited to feed-back corrections on the laser dc current.  The second servo loop is dedicated to steer the local oscillator frequency to the CPT resonance peak. The third one is devoted to compensate for light-shifts experienced by the atoms during the optical pulses.

\section{SABR sequence and electronics}
The clock experiment is fully controlled by a single FPGA board (Red Pitaya StemLab 125-14). The FPGA generates the SABR sequence shown in Fig. \ref{fig:1}(b). The sequence consists of two consecutive ABR sequences (ABR1 and ABR2), each composed of four two-step optical pulses of total length $T_b$. The use of two-step pulses when directly modulating a VCSEL has been shown effectively reducing the time required for the laser to reach the target atomic optical frequency \cite{Ide:IFCS:2015, Rivera:APL:2024}. The detection window, used for extracting the atomic signal, is opened for a duration $\tau_D$~=~1~$\mu$s, after a short delay $\tau_d$ of 18~$\mu$s from the beginning of the pulse. For each ABR sequence, the two first Ramsey-CPT patterns use a short dark time $T_S$ while the two next ones use a long dark time $T_L$. A $\pi/2$ phase jump is applied to the 9.192~GHz interrogation signal during the dark times such that the signal is respectively measured on both sides of the central Ramsey-CPT fringe. As indicated in Fig. \ref{fig:1}(b), $\pi/2$ phase jumps are here applied by generating frequency jumps $\Delta \nu_c \sim$~2.08~kHz, during a length $\Delta t$~=~60~$\mu$s. Error signals, noted $\varepsilon_S$ and $\varepsilon_L$, are then extracted for the short and long dark time patterns, respectively. In the second ABR sequence, a similar pattern is applied, except that the LO phase modulation pattern is inverted. This symmetric interrogation was demonstrated to be of significant importance in vapor cell clocks for canceling a memory effect of the atoms and subsequently improve the light-shift rejection efficiency \cite{MAH:APL:2018, MAH:2022}. Similarly to these previous studies \cite{MAH:APL:2018, MAH:2022}, two error signals are then ultimately generated. The first one, $\varepsilon_+$ = $\varepsilon_S + \varepsilon_L$, is used for stabilization of the LO frequency to the atomic transition. The second one, $\varepsilon_-$~=~$\varepsilon_S - \varepsilon_L$, is exploited for light-shift compensation.

Figure \ref{fig:2} presents a diagram illustrating the implementation of the sequence generation and data processing on the FPGA board. The system is specifically designed to enable precise control of the two servo loops described earlier. It operates at a 125 MHz clock frequency, provided by the analog-to-digital converter (ADC) on the Red Pitaya. The two-step sequence is stored in a Lookup Table (LUT), and its content is sent to a digital-to-analog converter (DAC 1), which is connected to the laser controller. The data acquisition process is synchronized with the start of the sequence generation. As shown in Fig.~\ref{fig:2}(a), the \mbox{``sequence start trigger"} signal, along with the ADC readings, is fed into the input of the ``Error Signal module".

Within the ``Error Signal module" (Fig. \ref{fig:2}(b)), the first block, ``Extract region", isolates a 2 $\mu$s segment at the beginning of each cycle, ensuring that only the portion of the signal corresponding to the correct wavelength is later processed. After the region of interest is extracted, the next stage is the ``Bottom tracking module", which returns a 1 $\mu$s segment centered around the bottom of the absorption profile. This step mitigates the effects of jitter in the position of the profile (see section \ref{sec:detwintrack}). The output of this module is then passed to the ``Sum of Frame" block, where it is averaged to produce a single representative point for each cycle. The averaged point is subsequently sent to the ``Central Difference" block, which computes the difference between successive points. This step leverages the modulation of the LO (based on the LO LUT content) to extract signals from the left and right sides of the Ramsey fringe. Finally, the ``Element Selector" ensures that only differences between consecutive cycles of the same type ($T_L$ or $T_S$) are used. This is essential due to the interleaved nature of the ABR sequence and the pipelined data processing implemented in the system.

The resulting error signals, $\varepsilon_S$ and $\varepsilon_L$, are used to generate error signals $\varepsilon_+$ and $\varepsilon_-$. These error signals are fed into their corresponding PID controllers for either frequency or phase correction. The corrections are applied to the LO sequence stored in ``LUT synth", either by adjusting its offset for frequency correction or modifying the amplitude ratio between the $\pm$2.08 kHz frequency jumps ($\pm\pi/2$ phase jumps) for phase correction. Finally, the modified LO sequence is sent to ``DAC 2", which is connected to the modulation input of the microwave synthesizer. Additionally, the system provides the flexibility to apply a ramp instead of frequency or phase corrections, allowing the observation of error signals. 

\begin{figure}[t!]
    \centering
    \includegraphics[width=1.0\linewidth]{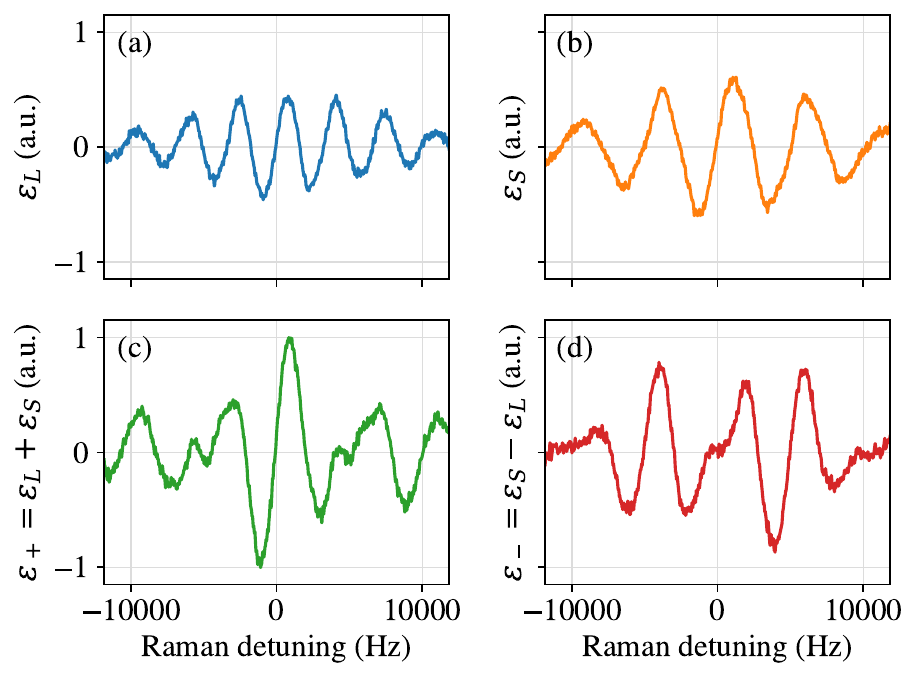}
    \caption{Error signals $\varepsilon_L$ (a), $\varepsilon_S$ (b), $\varepsilon_+$ (c) and $\varepsilon_-$ (d) extracted for a typical interrogation sequence with $T_b$~=~170~$\mu$s, $T_S$~=~150~$\mu$s and $T_L$~=~250~$\mu$s. $\varepsilon_+$ is used for the LO frequency stabilization while $\varepsilon_-$ (d) is exploited for light-shift compensation. The laser power $P_l$ at the cell input is 188~$\mu$W. Each error signal consists of 2046 points, and an 8-sample window moving average was applied to better highlight the signal features.}
    \label{fig:3}
\end{figure}
\begin{figure}[t!]
    \centering
    \includegraphics[width=1.0\linewidth]{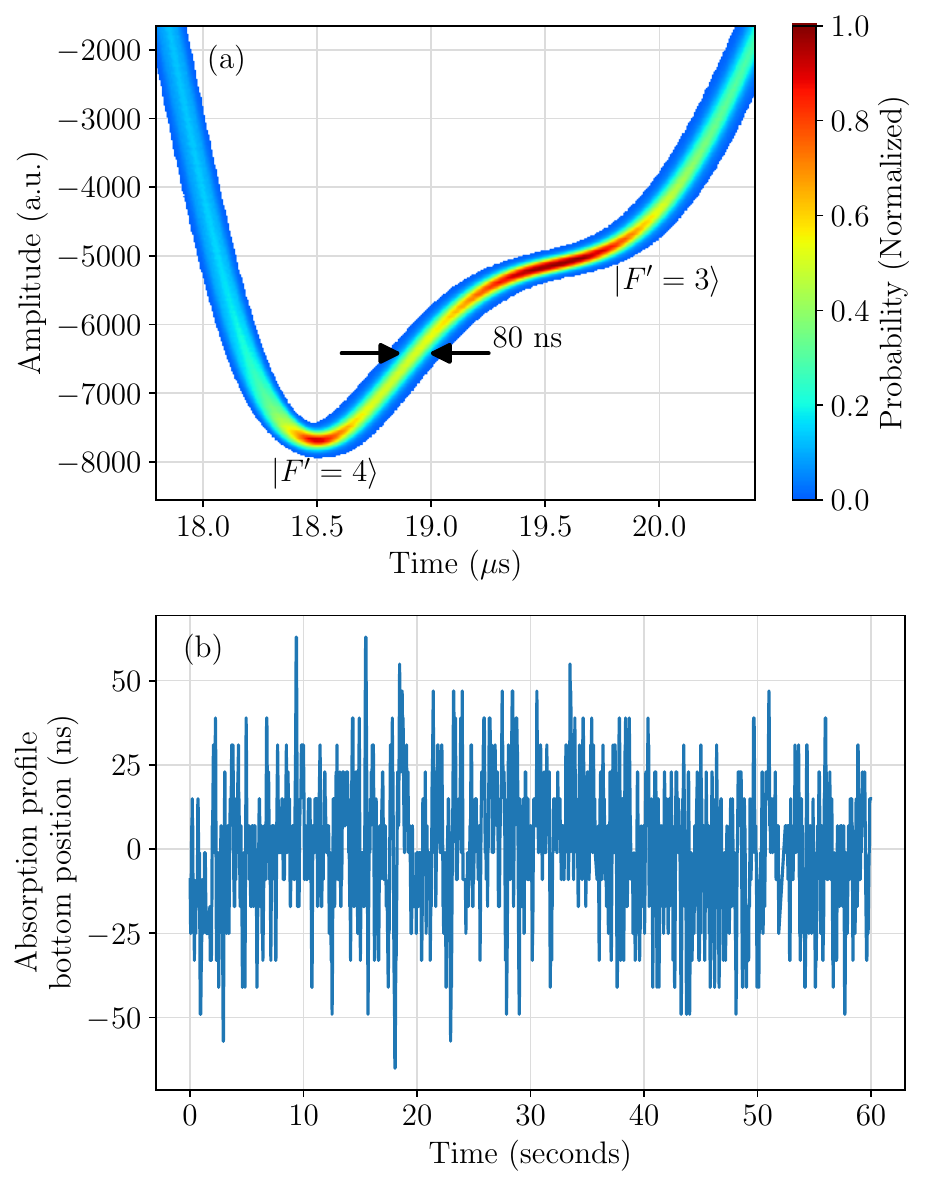}
    \caption{(a) Timing jitter of the absorption profile, and then of the detection window content, induced by current-actuated power modulation of the VCSEL. (b) Temporal trace of the absorption profile bottom position. The amplitude of its fluctuations is about 80~ns.}
    \label{fig:4}
\end{figure}

Figure \ref{fig:3} shows error signals $\varepsilon_L$ (a), $\varepsilon_S$ (b), $\varepsilon_+$ (c) and $\varepsilon_-$ (d) extracted for a typical interrogation sequence with $T_b$~=~170~$\mu$s, $T_S$~=~150~$\mu$s and $T_L$~=~250~$\mu$s. Pulses of duration $T_b$ are composed of a first step of length $\tau_1$~=~20~$\mu$s and a second step of length $\tau_2$~=~150~$\mu$s. The error signal $\varepsilon_S$, obtained in cycles with short dark time $T_S$, is broader and higher in amplitude than the error signal $\varepsilon_L$ obtained in cycles with long dark time $T_L$. The error signal $\varepsilon_+$ benefits from an enhanced amplitude that justifies its use for LO frequency stabilization. The error signal $\varepsilon_-$ shows in open-loop configuration a zero-crossing point, image of the light-shift to be compensated.
\begin{figure}[t!]
    \centering
    \includegraphics[width=\linewidth]{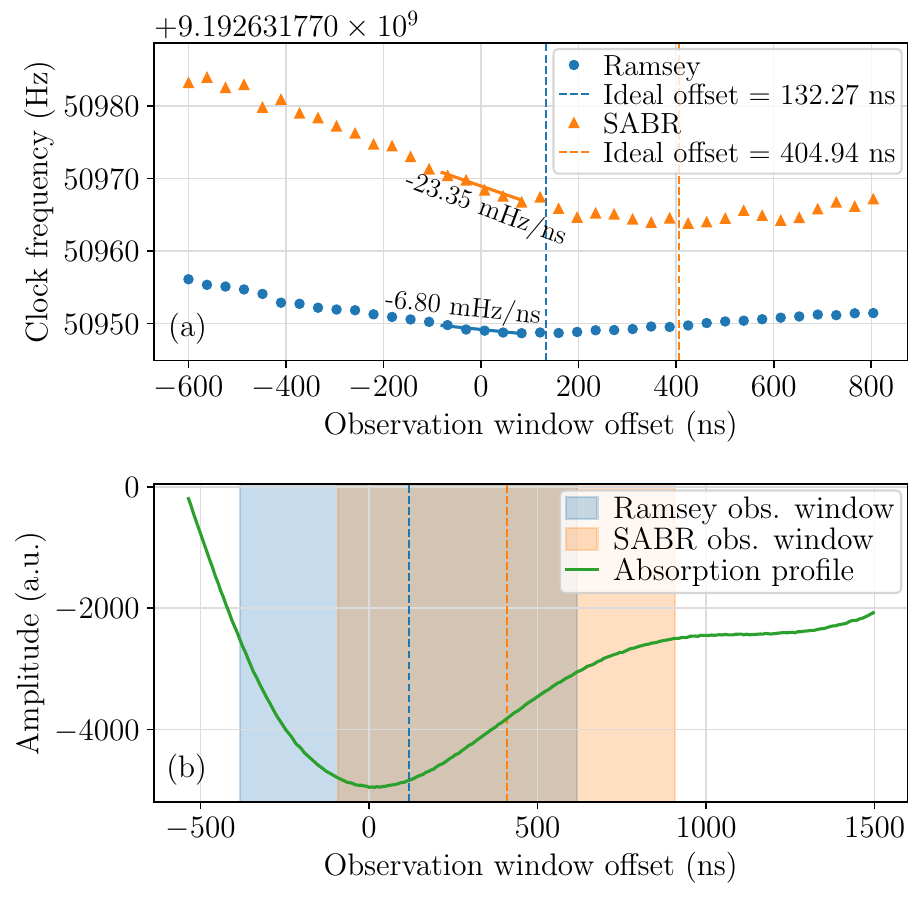}
    \caption{Evolution of the clock frequency with the observation window offset, in both Ramsey-CPT and SABR-CPT modes. In the two regimes, an offset set point range where a change of sign of the dependence is observed, is identified. Parameters are: $P_l$ = 188 $\mu$W and $P_{\mu W}$ = $-$5.5~dBm.}
    \label{fig:obs_offset}
\end{figure}

\section{Detection window tracking}
\label{sec:detwintrack}

Once the setup implemented and the FPGA programmed, our initial motivation was to measure light-shift coefficients, to assess the efficiency of the SABR-CPT sequence when applied through direct current-based power modulation. However, during initial tests, these measurements, particularly with respect to the laser detuning and the microwave power (which were not investigated in \cite{Rivera:APL:2024}), 
proved unreliable and unvalid, as the absorption line could move out of the detection window. In the present experiment, the abrupt current pulses induce transient changes of the VCSEL temperature, causing jitter of the laser frequency and consequently shifts of the absorption profile position. Since the atomic signal is extracted from the bottom of this profile, such shifts undermine measurement accuracy.


Figure \ref{fig:4}(a) illustrates the jitter of the absorption profile within the detection window, highlighting its temporal fluctuations. The signal was acquired within the FPGA system embedded in the Red Pitaya, directly at the output of the ``extract region" block in Fig.~\ref{fig:2}(b). A total of 36 180 absorption profiles were recorded, \mbox{non-consecutively}, due to data transfer bandwidth limitations, equivalent to 11.57~s of continuous acquisition. These profiles were compiled into a histogram to analyze the probability distribution of their positions over time, revealing fluctuations of approximately $\pm$40~ns around the position reference.

To observe the temporal evolution of this fluctuation, the position of the absorption profile minimum was extracted using the ``find bottom" block in Fig. \ref{fig:2}(c) and recorded continuously over a 60-s period, as shown in \mbox{Fig. \ref{fig:4}(b)}. This confirms that the absorption profile position exhibits significant temporal instability, consistent with observations reported in \cite{Fukuoka:JJAP:2023}.

We suspected that these fluctuations might affect the clock frequency. A detection window position sweep test was therefore implemented. This study aimed to identify an optimal operating point, and to extract a sensitivity coefficient. Figure \ref{fig:obs_offset}(a) shows the evolution of the clock frequency versus a timing offset, applied to the 1~$\mu$s-long observation window, relative to the position of the absorption profile bottom at 18.5 $\mu$s from the start of the pulse, in both Ramsey-CPT and SABR-CPT modes. In the Ramsey-CPT mode (with $T_b$~=~170~$\mu$s and $T$~=~150~$\mu$s), we observe a turnover offset point ($\sim$~132~ns), where the clock frequency dependence is canceled at the first order. Around the null offset point (0 of the x-axis), the sensitivity of the clock frequency to the detection window timing is $-$6.8 mHz/ns, i.e. 7.4$\times$10$^{-13}$/ns relative to the clock frequency. In the SABR-CPT mode, a stronger dispersion of frequency data points explained by the degradation of the clock short-term stability is observed. Nevertheless, in this mode, we find a wide region (offset between 200 and 600 ns) in which the clock frequency does not change much. Around the null offset point, the sensitivity of the clock frequency to the detection window timing is in the SABR-CPT mode $-$23.35 mHz/ns, i.e. $-$2.5$\times$10$^{-12}$/ns in fractional value. The ideal offset points, shown as dashed lines in Fig.~\ref{fig:obs_offset}(a), are illustrated in Fig.~\ref{fig:obs_offset}(b) in contrast to the absorption profile. These results emphasized the importance of precisely compensating for the detection window jitter.

To solve this issue, we have implemented on the FPGA a custom real-time absorption line position tracking functionality. This system, depicted in Fig. \ref{fig:2}(c), detects the absorption line bottom position within a 1~$\mu$s window and then extracts the data around it. This functionality revolves around the ``Extract around bottom" block, which processes a stream of samples (left input as depicted in the diagram) and store them in memory, with each sample assigned an index in an array. Simultaneously, the block receives an input (on top in the diagram) that indicates the location of the bottom. Once a 2 $\mu$s window of samples has been collected (providing sufficient data selection), the algorithm calculates the necessary offset to the element indices to ensure that the central element of the produced data corresponds to the detected bottom. Finally, a sequential writing process generates a 1~$\mu$s-long dataset centered around the identified bottom. Note that it is also possible to intentionally apply an arbitrary offset to the output of the "find bottom" block to enable the extraction of data around an ``ideal" offset for each mode. The implemented method allows the real-time tracking and correction for jitter on every single Ramsey-CPT cycle.

In the SABR-CPT regime, we found that the clock stability could not average down 
after a few hundreds of seconds before implementation of the detection window tracking. This technique has allowed to significantly mitigate the detrimental impact of the observation window jitter on the clock stability, and has revealed to be a crucial step for performing reliable and repeatable light-shift measurements, as well as demonstrating a clock with enhanced mid- and long-term stability.

\section{Light-shifts}
\begin{figure}[h!]
    \centering      \includegraphics[width=0.98\linewidth]{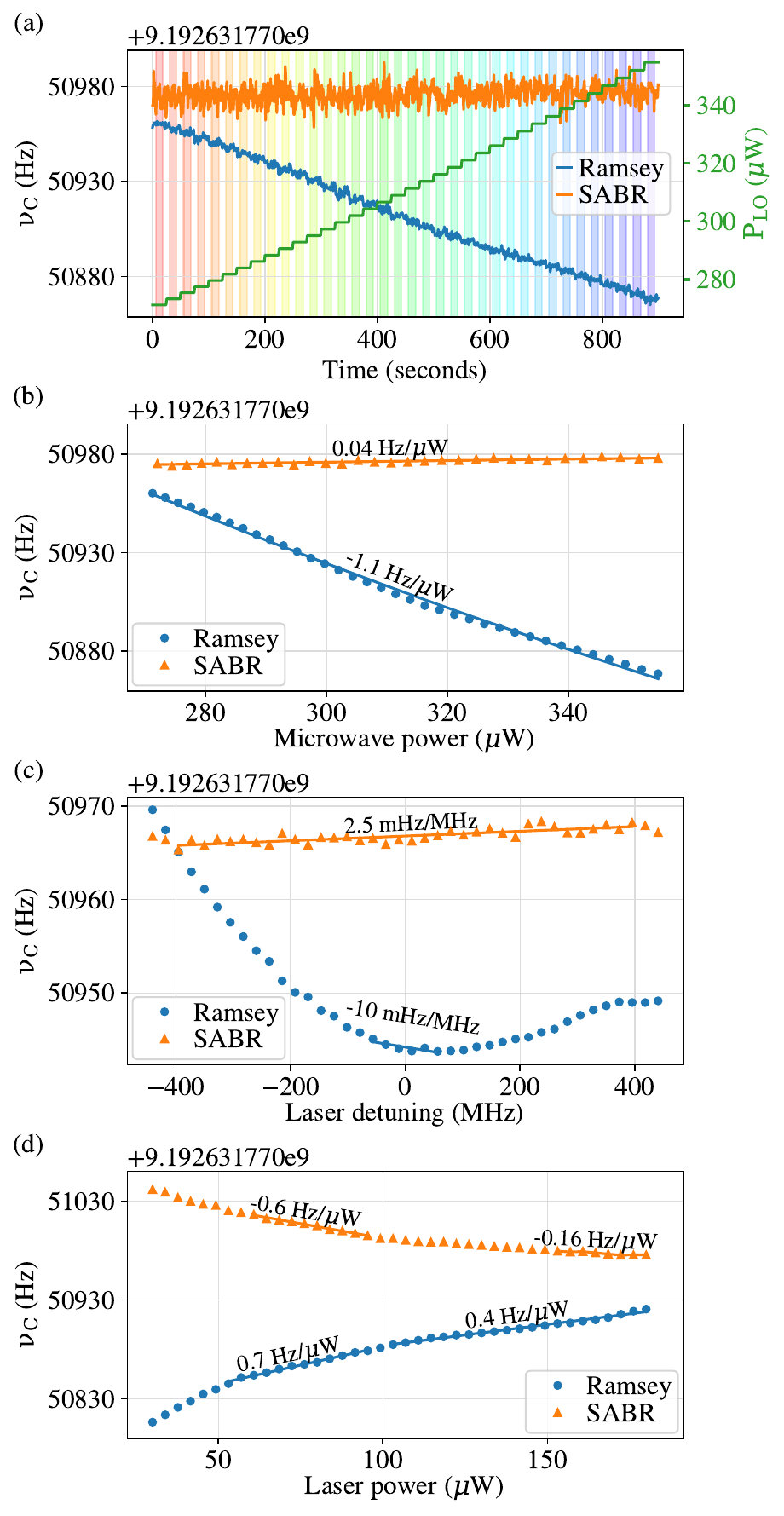}
    \caption{(a) Temporal sequence of the clock frequency. Each step results from the change of the microwave power, inducing a light-shift. Ramsey-CPT and SABR-CPT cases are compared. In the SABR-CPT regime, the clock frequency remains about unchanged despite microwave power jumps. For each microwave power step, the average value of the clock frequency is extracted. Derived from such a sequence, the sub-plot (b) shows the frequency shift of the clock frequency, relative to the unperturbed Cs atom frequency (9.192 631 770~GHz) versus the microwave power. Following a comparable procedure, sub-plots (c) and (d) report the clock frequency shift versus the laser frequency and the laser power, respectively. To change the laser power, a variable neutral density filter was placed before the microfabricated vapor cell (not shown in Fig. \ref{fig:1}). If not varied, we use $P_l$~=~188 $\mu$W and $P_{\mu W}$~=~$-$5.5~dBm (282~$\mu$W).}
    \label{fig:6}
\end{figure}
\begin{figure*}[t!]
    \centering
    \includegraphics[width=\linewidth]{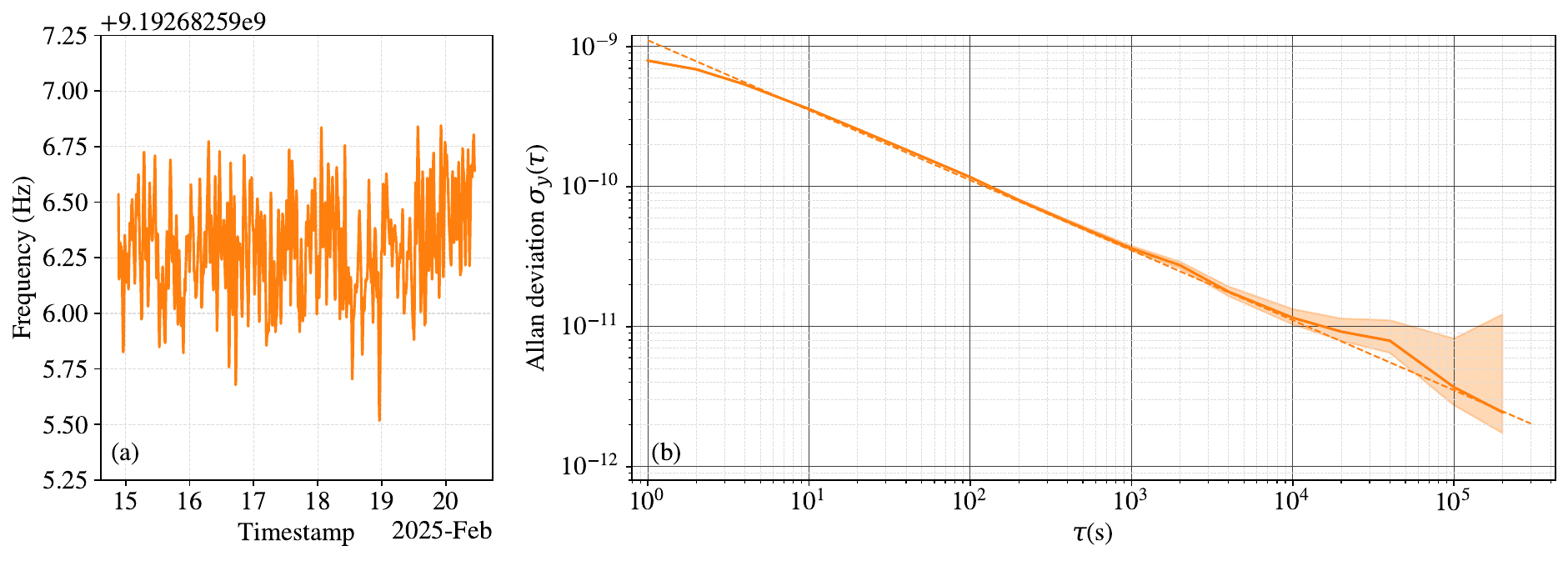}
    \caption{(a) Temporal trace of the clock frequency over 5 days. A 1-hour moving average is applied for illustrative purposes only. (b) Allan deviation of the clock frequency over the 5-day period.
    The dashed orange 
    line represents the slope of $1.1\times10^{-9} \tau^{-1/2}$. 
    }
    \label{fig:7}
\end{figure*}

Figure \ref{fig:6}(a) shows temporal traces of the clock frequency in Ramsey-CPT (in blue, $T$~=~150~$\mu$s) and SABR-CPT (in orange, $T_S$~=~150~$\mu$s, $T_L$~=~250~$\mu$s) modes. While the clock is running, light-shifts are willingly applied by changing the microwave power that drives the VCSEL. In the Ramsey-CPT case, the clock frequency changes abruptly every time the microwave power is changed. On the other hand, in the SABR-CPT mode, the clock frequency, despite a visible degradation of the signal-to-noise ratio, remains nearly constant. Figure \ref{fig:6}(b) resumes the clock frequency shift versus the microwave power in both Ramsey-CPT and SABR-CPT cases. In the Ramsey-CPT case, we measure a sensitivity of $-$1.1~Hz/$\mu$W. In the SABR-CPT case, we observe a reduction of the clock frequency variations by a factor of 27.5, yielding the coefficient of $+$0.04~Hz/$\mu$W. This value is close from the one obtained in \cite{MAH:2022}, where the pulsed sequence was produced with an AOM.

Figures \ref{fig:6}(c) and \ref{fig:6}(d) report comparable studies as a function of the laser frequency and the total laser power, respectively. Clock frequency sensitivity tests to laser frequency variations were performed by sweeping the laser detuning over a wide range of about $\pm$~450~MHz. In this range, the clock frequency changes in the Ramsey-CPT case by almost 30 Hz. A sign change of the slope is found close to the null detuning. In the SABR-CPT case, the response is clearly flatter within the tested range, yielding an extracted linear coefficient of about 2.5~mHz/MHz, or 2.7~$\times$~10$^{-13}$ in fractional value. This value is comparable to the one obtained in \cite{MAH:2022}. Interestingly, we found in this test (Fig. \ref{fig:6}(c)), by running the clock in the SABR-CPT regime, that phase corrections applied to the LO to compensate for the light-shift had the same trend than the light-shift curve obtained in the Ramsey-CPT case. This confirms the efficiency of the SABR sequence to tackle residual laser frequency-induced shifts observed in the Ramsey-CPT mode.

For the laser power (Fig. \ref{fig:6}(d)), the light-shift trends in both modes exhibit opposite signs (negative for SABR and positive for Ramsey-CPT). In the SABR-CPT case, the light-shift slope is found to be reduced with increased laser power. This behaviour was also observed in a high-performance SABR-CPT clock \cite{MAH:APL:2018}, motivating the preference to operate the clock at high laser power to mitigate the light-shift coefficient. Specifically, in the tested range here, the light-shift slope in the SABR-CPT mode becomes significantly reduced for $P_l >$ 160~$\mu$W. In the Ramsey-CPT case, the light-shift tends to increase linearly for power values higher than about 120 $\mu$W. This suggests that the gain offered by SABR in light-shift mitigation, with respect to the Ramsey-CPT case, might be enhanced at even higher laser power values. With the SABR-CPT sequence, the sensitivity, obtained by fitting the last 8 data points of the curve, is measured to be $-$0.16~Hz/$\mu$W, a factor about 2.5 times lower than the one obtained in the Ramsey-CPT case. 

\section{Frequency stability}
\label{sec:Stability}
In the final part of the study, we evaluated the fractional frequency stability of the microcell CPT clock in the SABR-CPT case ($T_S$~=~150~$\mu$s, $T_L$~=~250~$\mu$s, $T_b$~=~170~$\mu$s, $\tau_d$~=~18~$\mu$s and $\tau_D$~=~1~$\mu$s). A 5-day measurement was conducted using our table-top microcell CPT clock prototype. No active laser power or microwave power servo are implemented.

Figure \ref{fig:7}~(a) shows the temporal trace of the clock frequency while Fig. \ref{fig:7}(b) reports the corresponding Allan deviation. The clock fractional frequency stability is 1.1~$\times$~10$^{-9}$ at 1~s and averages down, with a $\tau^{-1/2}$ slope, signature of white frequency noise, to reach the level of 2.4~$\times$~10$^{-12}$ at 2~$\times$10$^5$~s.

We have observed that switching from the Ramsey-CPT to the SABR-CPT sequence (without window tracking) induced a degradation by a factor of about 4.3 on the short-term stability. We have also noted that the activation of the detection window tracking technique could induce an excess degradation by a factor of about 1.9. We think that the short-term stability of the clock in the SABR-CPT mode could be improved, at least by a factor 2, through some optimizations of the clock parameters and by adopting the laser amplitude noise normalization technique implemented in \cite{MAH:APL:2018}. Efforts are in progress in this direction.

The stability level of 3.7~$\times$~10$^{-12}$ at 1~day is comparable to the one obtained in \cite{Carle_OE_2023}, in which the pulsed sequence was produced with an external AOM, and in cases where no laser power and microwave power servos were applied. Results obtained here using laser current-actuated power modulation of the VCSEL are then very encouraging and approach performances of best reported CPT-based CSACs \cite{Yanagimachi:APL:2020, ULPAC:2019, Carle_OE_2023}. Before concluding, we note that the clock laser has been now operated in the pulsed regime for more than 18 months. No visible signs of laser or clock signal degradation have been observed so far.

\section{Conclusions}
In conclusion, we have demonstrated a CPT-based microcell atomic clock using the pulsed SABR sequence. The SABR sequence was implemented by directly modulating the dc current of the VCSEL, avoiding the use of any external AOM or optical shutter. The sequence generation, data acquisition, processing and corrections for servo loops are handled by a single Red Pitaya FPGA-board. To ensure proper clock operation and mitigate limitations caused by current-induced thermal transients in the laser, a real-time detection window tracking system was implemented. The clock frequency dependence to microwave power, laser power and laser frequency variations was significantly reduced with respect to Ramsey-CPT spectroscopy, up to a factor 27.5 for the microwave power. Over a 5-days measurement, the SABR-CPT microcell clock demonstrates a fractional frequency stability of 1.1~$\times$~10$^{-9}$ at 1~s and 2.4~$\times$~10$^{-12}$ at 2~$\times$~10$^5$s. The stability at 1~day is comparable to the one obtained in \cite{Carle_OE_2023} (without laser and microwave power servos) and among best CPT CSACs reported in the literature so far \cite{Yanagimachi:APL:2020, ULPAC:2019, Carle_OE_2023}.

\section*{Acknowledgment}
This work was supported by the Direction G\'{e}n\'{e}rale de l’Armement (DGA) and by the Agence Nationale de la Recherche (ANR) in the frame of the ASTRID project named PULSACION (Grant ANR-19-ASTR-0013-01), LabeX FIRST-TF (Grant ANR 10-LABX-48-01), EquipX Oscillator-IMP (Grant ANR 11-EQPX-0033) and EIPHI Graduate school (Grant ANR-17-EURE-0002). The PhD thesis of C. Rivera is co-funded by the program FRANCE2030 QuanTEdu (Grant ANR-22-CMAS-0001) and Centre National d'Etudes Spatiales (CNES). This work was partly supported by the french RENATECH network and its FEMTO-ST technological facility (MIMENTO).

\section*{Data availability statement}
The data of this study are available from the corresponding author upon reasonable request.

\section*{Conflict of interest}
The authors state that there is no conflict of interest to disclose.




\providecommand{\noopsort}[1]{}\providecommand{\singleletter}[1]{#1}%

\end{document}